\documentclass[onecolumn, numberedappendix]{aastex702}
\usepackage{graphicx}%{color}
\usepackage{color}

\newcommand{\av}{$A_V$}

\newcommand{\eg}{{\it e.g.}}

\newcommand{\ks}{$K_{\rm s}$}
\newcommand{\vmkz}{$(V-K_{\rm s})_0$}
\newcommand{\vmk}{$(V-K_{\rm s})$}

\newcommand{\mum}{$\mu$m}

\begin{document}

\title{Rotation in the Lagoon Nebula with K2}

\reportnum{Version from \today}

\author[0000-0001-6381-515X]{L.~M.~Rebull}
\affiliation{Infrared Science Archive (IRSA), IPAC, MS 100-22, 1200 E.\
California Blvd., California Institute of Technology, Pasadena, CA
91125, USA}
\email{rebull@ipac.caltech.edu}

\author[0000-0002-3656-6706]{A.~M.~Cody}
\affiliation{SETI Institute, 339 Bernardo Ave., Suite 200, Mountain View, CA 94043, USA}
\email{acody@seti.org}

\author[0000-0002-4115-0318]{L.~Venuti}
\affiliation{SETI Institute, 339 Bernardo Ave., Suite 200, Mountain View, CA 94043, USA} 
\affiliation{Visiting Fellow, School of Physics, UNSW Science, Kensington, NSW 2052, Australia}
\email{lvenuti@seti.org}

\author[0009-0005-8540-0963]{S.~Thanvantri}
\affiliation{Department of Computer Science, Princeton University, 35 Olden St, Princeton, NJ 08544, USA} 
\email{thanvantri@princeton.edu}

\author[]{A.~Thanvantri}
\affiliation{Harvard John A. Paulson School of Engineering and Applied Sciences, 150 Western Ave, Allston, MA 02134 USA} 
\email{apoorva.thanvantri@gmail.com}

\begin{abstract}
A low-mass star's rotation rate drives its magnetic field, and in the
case of young stars, also affects accretion and planetary formation.
Observations of the rotation rates of young stars help constrain models
of star and planet formation. 
We present rotation rates from K2 observations of stars in the 
$\sim$1-2 Myr Lagoon Nebula, one of the youngest clusters with
space-based rotation rates available.  From 414 light curves we obtain
reliable rotation periods for 325 stars; 124 of them are disk-bearing
and 201 are disk-free. This constitutes  a significant increase in the
collection of K2 rotation rates  available for $\sim$1-5 Myr stars.
The overall distribution is  broadly consistent with expectations, in
that the slowest rotators are mid-K and both lower and higher masses
are on average faster rotating.  There is weak evidence that disked
stars rotate more slowly. There are a few complex periodic variables,
but they appear to be surprisingly high mass compared to prior known
examples of this kind of object.
\end{abstract}
%\keywords{tocome}

\section{Introduction}
\label{sec:intro}%\textcolor{red}{sec:intro}

The evolution of stellar rotation rates during the first few million
years of a star's life is important for understanding how -- and
whether -- stars form planets. For young low-mass stars, cool
magnetospheric starspots rotating into and out of view create changes
in the measured light from the star, variations that are large and
easy to measure compared to planetary transits. For stars that still
have circumstellar disks, the interaction between the star and its
disk as the young stellar object (YSO) contracts onto the main
sequence is revealed via brightness fluctuations arising from
photospheric cool spots or accretion hot spots rotating into and out
of view. Non-uniformities in the disk or accretion columns can occult
the central YSO from our perspective, and changes in the accretion
rate can all result in changes in the light we detect.  If we can
determine the stellar rotation rate despite these flickering
contributions, conservation of angular momentum through these early
phases of a star's life can be revealed in the evolution of the
stellar rotation rate through disk dispersal and possibly planet
formation. The variations of YSOs are easily detected by missions
designed for planet hunting, like NASA's K2 \citep{k2howell}.

Studying stellar rotation rates in the youngest clusters is complicated 
not only because of the often stochastic contribution from disks and
accretion, but because of the dusty natal environments. Reddening affects
the measured colors and brightnesses, which affects the inferred 
distances, temperatures, masses, and ages. Rotation periods are 
often the quantity known with the highest accuracy and precision
in young stars. 
%Rotation rate distributions for these youngest clusters are often hard-fought victories.

\begin{figure}[htb!]
\epsscale{1.0}
\plotone{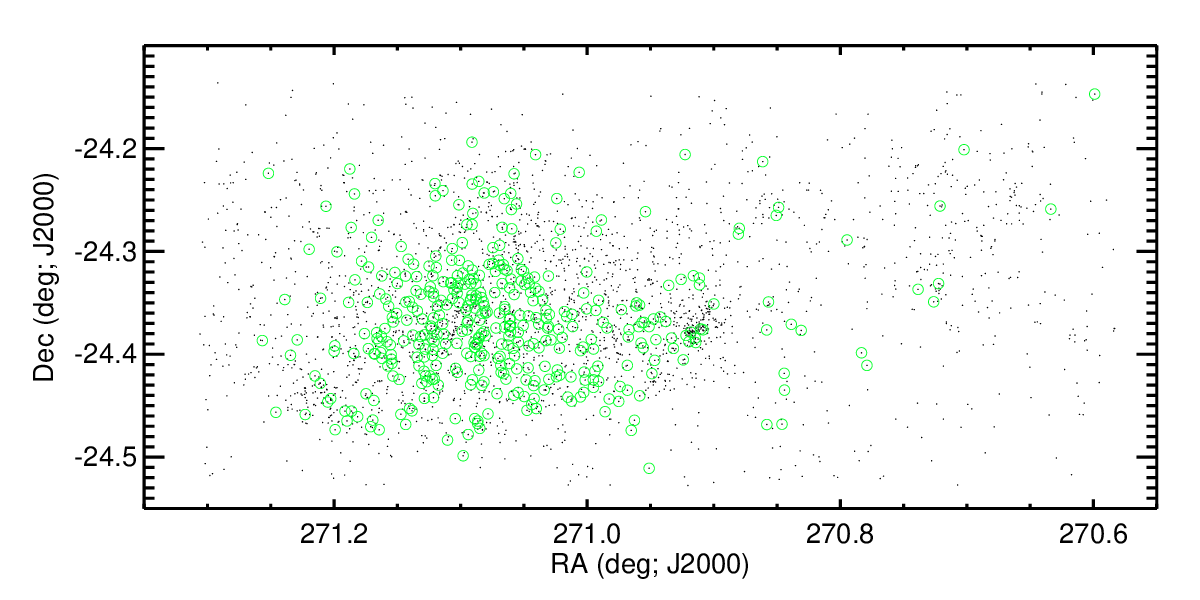}
\caption{A plot on the sky of the YSO candidates in the master catalog from 
\citet{lagooncatalog} (black dots) with the targets from this paper (those
with K2 light curves) indicated via an additional green circle.}
\label{fig:where}
\end{figure}

The Lagoon Nebula is a high-mass star-forming region located in projection
near the Galactic Center.  NGC 6530 is an open cluster slightly in front 
of the nebulosity that is also part of M8.
Stars in M8 are known to be young 
\citep[$\sim$1-2 Myr;][]{2019AA...623A.159P}, but it is relatively far away  
at $\sim$1250 pc \citep{2019AA...623A.159P}.  
In \citet{venuti2021,venuti2024}, we have discussed results from multicolor
ground- and space-based monitoring of the Lagoon Nebula, and explored
the influence of environment and mass on circumstellar disk accretion.
In \citet{lagooncatalog}, we presented a catalog of literature-selected
YSOs optimized for this region and assembled SEDs from archival 
photometric data. The work in the current paper depends on the 
catalog from \citet{lagooncatalog}, relying on the analysis there for
SEDs, disk identification, and reddening estimates. Figure~\ref{fig:where} 
shows the location on the sky of the YSOs from this catalog.

In this paper, we use the K2 observations of the Lagoon Nebula 
originally presented in \citet{venuti2021} to infer rotation rates for
the YSOs. There are 414 unique light curves (these objects are
highlighted in Fig.~\ref{fig:where}), and the Lagoon is one of the
youngest clusters for which there are substantial numbers of
space-based rotation rates now available. The confusion arising from
the surface density of sources and the stochastic contributions to the
light curves from accretion and circumstellar disks both make this
analysis difficult.

We begin in Sec.~\ref{sec:obsanddata} by summarizing the ancillary
data obtained from \citet{lagooncatalog} and the data directly
relevant to the analysis in this paper; we also summarize the K2 data
reduction and our period-finding approach in this section. 
Sec.~\ref{sec:pdistrib} discusses the distribution of $P$ vs. \vmkz\ 
for the Lagoon on its own and in context with other clusters.  We also
discuss the influence of disks. Sec.\ref{sec:other} highlights a few 
of the other kinds of light curves, including particularly the complex
periodic variables identified here.  Then we summarize the results in
Sec~\ref{sec:summary}.

\section{Observations and Data Reduction}
\label{sec:obsanddata}

\subsection{Ancillary Data}

We assembled a large catalog of literature YSOs in
\citet{lagooncatalog}, optimized for the region encompassed by the K2
data. There, we also identified the YSOs with disks, and resolved
source confusion issues within that catalog as best we could. The K2
pixels are large enough ($\sim4\arcsec$), and the source surface
density is high enough, that we used our master catalog (along with
selected image inspection) to identify light curves likely to be
subject to spatial confusion issues.

Three of the stars with K2 light curves (J180416.31-242503.1,
J180431.91-242326.1, and J180449.53-241522.0) are \citet{venuti2024}
members not otherwise identified as YSOs in the literature, and so do
not appear in the table in \citet{lagooncatalog}, though they were
treated the same way as the other YSOs and candidates there; their
properties are given in Table~\ref{tab:data}.

Since we ultimately needed $V$ magnitudes to put these stars in the
same rotation parameter space as similar data from other clusters, we
specifically sought measured $V$ from the literature. More than 90\%
of the stars with light curves have a measured $V$ magnitude. As can
be seen in \citet{lagooncatalog}, the SEDs are in general very well
defined in the optical. For the stars without measured $V$  in
\citet{lagooncatalog}, as in other of our papers
\citep[\eg,][]{ucllcc}, we interpolated a value for $V$ from a fit to
the SED. $JH$\ks\ measurements come from 
the Two Micron All-Sky Survey \citep[2MASS;][]{2masspaper,2masspsc}.

As discussed in \citet{lagooncatalog}, the reddening in this region
is inhomogenous and can be substantial.  If we have a spectral type
estimate, we can compare the observed \vmk, or the interpolated $V-$
the observed \ks, to the expected \vmk\ for the spectral type to get
an estimate of the reddening.

We have spectral type estimates for $\sim$90\% of the sample with 
light curves. For the
remaining stars, if we have $JHK_s$ for the stars (which we do for
most of the remaining catalog), we can place the stars on a $J-H$ vs.\
$H-K_s$ diagram, then shift them back along the reddening law derived
by \citet{Indebetouw} to the expected $JHK_s$ colors for young stars
from \citet{pecautmamajek} or the T~Tauri locus from
\citet{meyerttlocus}, and convert to $E(V-K_s)$ via $A_K = 0.114
A_V$ \citep{ccm}.  

\begin{figure}[htb!]
\epsscale{0.5}
\plotone{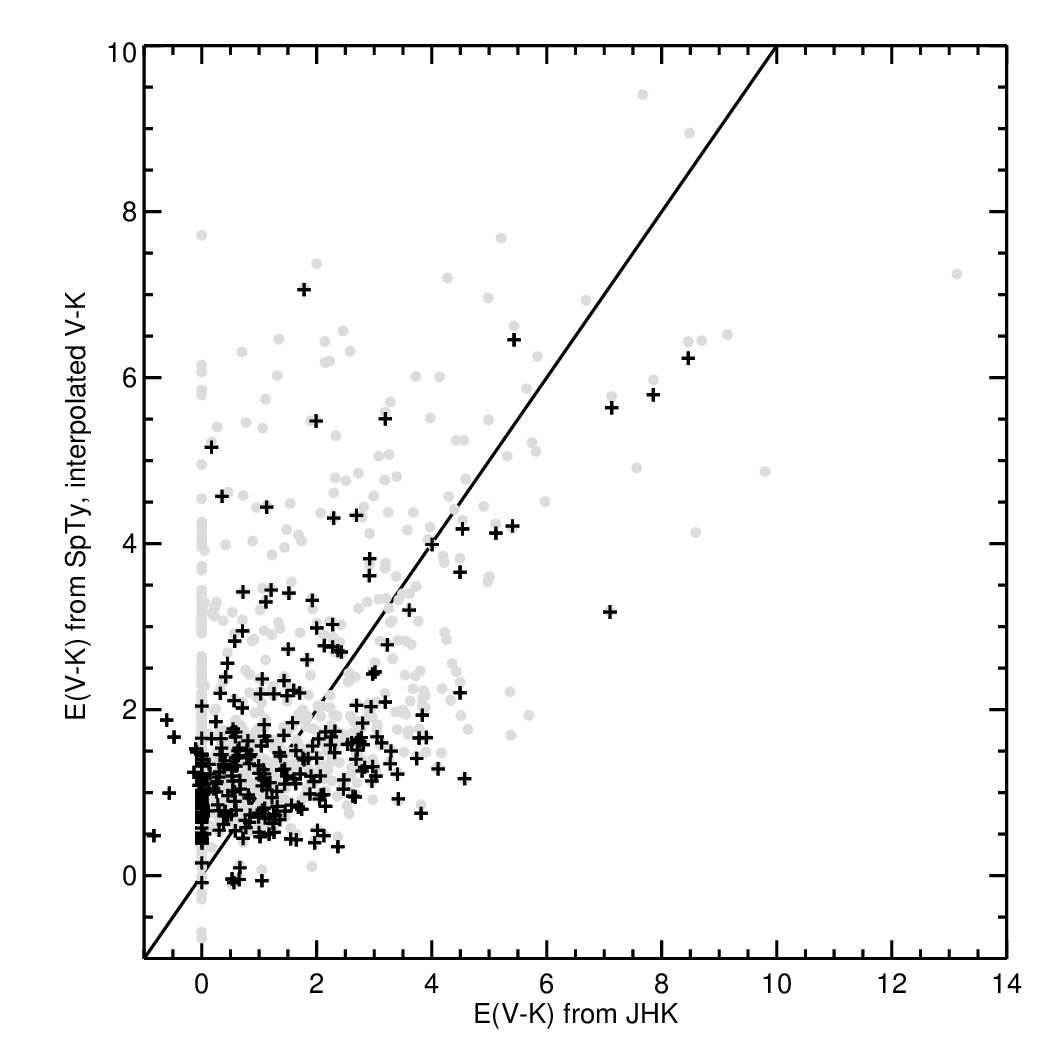}
\caption{Comparison of $E$\vmk\ for stars where the reddening is estimated
from the placement in the $JHK_s$ diagram ($x$-axis) and the $E$\vmk\
calculated from the observed \ks, interpolated $V$, and colors assumed  
from the spectral type.  A 1:1 line is overplotted; the grey underplotted
dots are the values from the entire catalog \citep{lagooncatalog}.
There is a weak correlation, but it displays less scatter than for 
the entire catalog from \citet{lagooncatalog}.}
\label{fig:comparejhksptyav}
\end{figure}

\begin{figure}[htb!]
%\epsscale{0.5}
\plotone{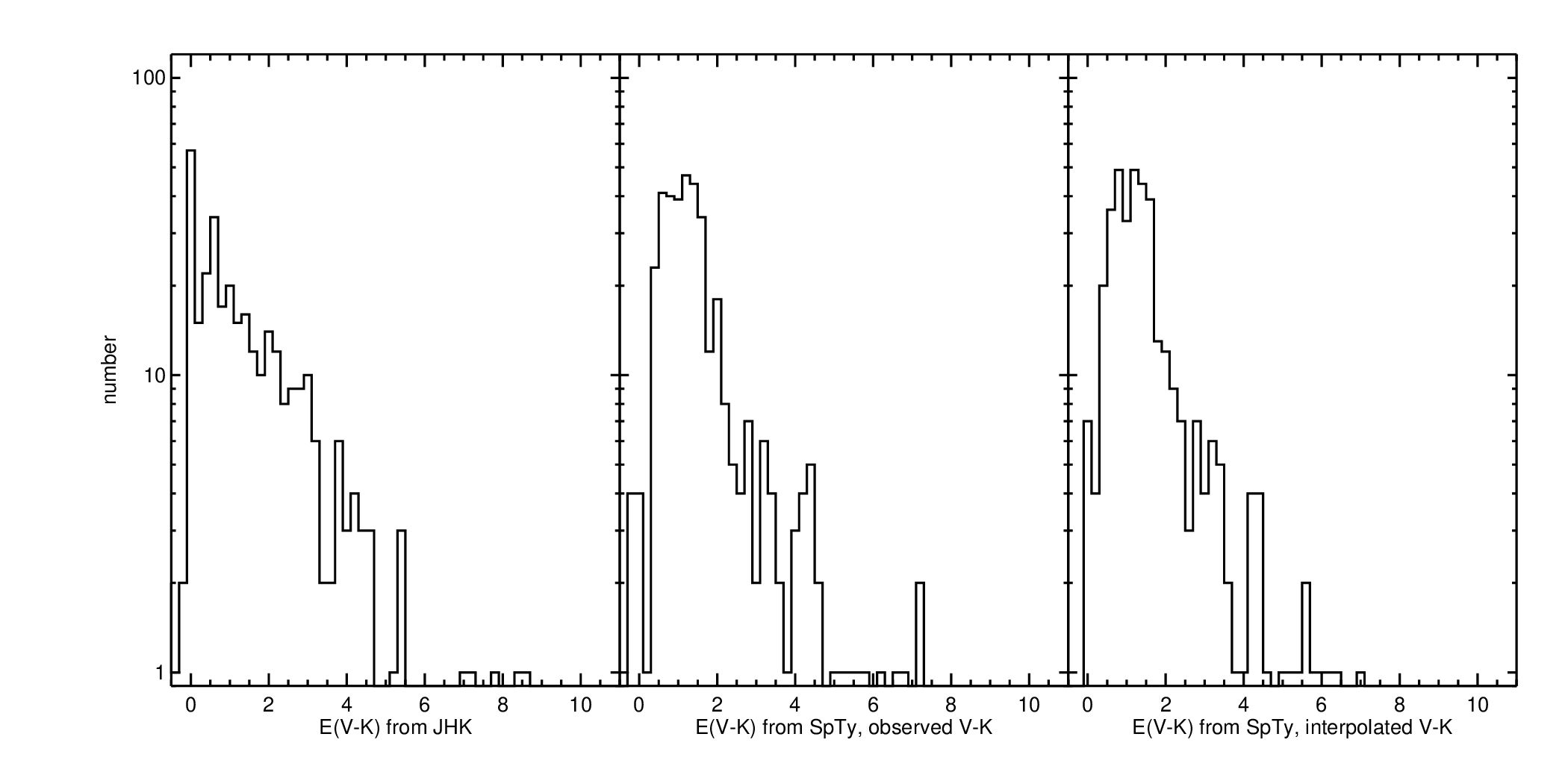}
\caption{Comparison of $E$\vmk\ for stars where the reddening is estimated
from the placement in the $JHK_s$ diagram (left), the $E$\vmk\
calculated from the observed $V$ and \ks\ (middle) and colors assumed  
from the spectral type, and 
the $E$\vmk\ calculated from the interpolated $V$ and colors assumed  
from the spectral type (right). See the text for more discussion.}
\label{fig:evkhistos}
\end{figure}

As discussed in \citet{lagooncatalog}, the reddening as calculated via
those two methods may be very different.
Figures~\ref{fig:comparejhksptyav} and \ref{fig:evkhistos} are similar
to those in \citet{lagooncatalog} but only include the subsample with
light curves.  The $E$\vmk\ as calculated from $JHK_s$ is somewhat
correlated with the $E$\vmk\ from spectral type, but the correlation
is not particularly strong. The $E$\vmk\ as calculated from $JHK_s$
seems to peak more or less near 0; however, as described in 
\citet{lagooncatalog}, the set of values at exactly 0 are affected 
by cases where the routine failed to converge on \av$>$0, or failed 
to converge at all. In contrast, when $E$\vmk\ is calculated using
spectral types, the distribution peaks at $\sim$1.25-1.3 mag, 
comparable to that for the entire catalog. There is a long tail to 
large extinction values, though not as long or large as the entire catalog in
\citet{lagooncatalog} -- this is expected, given that the criteria to
be included here include a viable K2 (optical) light curve, so very
embedded objects are excluded.

In the end, the only stars with nominal light curves for which we 
lack estimates of reddening are three for which we are missing
\ks\ measurements due to source confusion. Since 2MASS can't resolve 
these stars, certainly K2 can't either, and the K2 light curves 
are also negatively impacted by source confusion. These stars'
light curves were functionally identical to that from their 
apparent companion and were dropped from the analysis.

Table~\ref{tab:data} includes the best available \ks, the observed $V$
(when available), the best available \vmk\ and \vmkz, and the indication
of whether or not the star has a disk from \citet{lagooncatalog}.

\subsection{K2 Time Series}

The K2 observations and their data reduction are described in more
detail in  \citet{venuti2021}. In summary, K2 observed the Lagoon
nebula region from 2016 April 21 -- July 1. There were some individual
stars targeted with postage stamps, but most of the cluster was
monitored via a superstamp. The K2 campaign sampled every  $\sim$30
minutes for $\sim$70 days. 

Aperture photometry was extracted for the targets, with the best
possible aperture selected individually for each target. The light
curves we generated are available as part of the K2YSO high-level 
science product in the Mikulski Archive for Space Telescopes: 
\dataset[10.17909/44f5-5e13]{http://dx.doi.org/10.17909/44f5-5e13}
\citep[][]{k2ysodata,codyhill2018,venuti2021,codyhill22}. 

\subsection{ZTF Time Series}

Because the source surface density is so high and the K2 pixels are so
large, in some cases, source confusion posed an issue. The data from
the Zwicky Transient Facility
\citep[ZTF;][]{ztfbellm,ztfmasci,ztfdata} are high enough spatial
resolution that in many cases, it can resolve the sources, but it does
not have as high sensitivity or as rapid time sampling in comparison
to K2. ZTF typically  samples once per night or few nights, over 8
years, in $g$ and/or $r$-band, and the pipeline photometry typically
has scatter of at least a few percent.  Because the sampling is
nightly, there is substantial aliasing with 1 d, 0.5 d, etc.  ZTF
extracted light curves are housed at IRSA \citep{ztfdata}; we searched
by position to find the closest object to our target.

We searched for periods in these data using the same approach as we used
for the K2 data (see next subsection), but because of the wildly different 
sampling and sensitivity, we only used ZTF to confirm periods from K2, or to
allocate the period(s) from K2 appropriately to stars resolved in ZTF
but unresolved by K2.  These cases are noted in Table~\ref{tab:data} 
where relevant. In about
one-third of the stars with available ZTF light curves, corroborating 
evidence for the K2 period can be found in one or both of the available 
ZTF light curves.

Public ZTF data are not available for every target. In one case,
where there was no public ZTF data but it still seemed that ground
based data would be particularly helpful, we consulted ASAS-SN
\citep{asassn1,asassn2} to check on the K2 period.  This is noted in
Table~\ref{tab:data}.

\subsection{Measurement of Rotation Periods}
\label{sec:gettingp}

To identify significant periods, we used the Lomb-Scargle \citep{scargle82}
approach as implemented by the NASA Exoplanet Archive Periodogram
Service\footnote{https://exoplanetarchive.ipac.caltech.edu/cgi-bin/Pgram/nph-pgram}
\citep{akeson13}. We used the Infrared Science Archive (IRSA)
Time Series
Tool\footnote{http://irsa.ipac.caltech.edu/irsaviewer/timeseries},
which employs the same underlying code as the Exoplanet Archive
service, but allows for interactive period selection. We generally 
searched a period range from 0.05 days (1.2 hours) through 20 days. 
In some cases, we deliberately sought shorter periods where warranted.

For each significant peak, we phased the light curve and examined it, 
omitting outliers if need be to assess the light curve.  Through this
process, we identified and omitted factor-of-two aliases and harmonics; 
we identified and retained additional real periods.
We retained up to 4 periods for each star; $\sim$25\% of the periodic
light curves have at least two periods, but only $\sim$3\% of the sample have
at least four periods.

We classified the light curves according to the schemes we have
established in earlier papers \citep[\eg,][and references
therein]{ucllcc}.  These classifications are discussed in more detail
below and overlap with  those from \citet{venuti2021}.  In most cases,
the classifications between the two papers are consistent with each
other. In several cases, we had to reassess the classification with
respect to the one presented in \citet{venuti2021} -- for example, 
some of the stars that are listed there as not periodic, upon closer 
consideration here, are periodic. 

The periods we derive appear in Table~\ref{tab:data}, along with the 
light curve classifications.

\newcommand{\cellp}[1]{\parbox[t]{11cm}{\raggedright #1}}

\begin{deluxetable}{cl}%p{6cm}}
%\tabletypesize{\scriptsize}
%\rotate
\tablecaption{Contents of Table: K2 Rotation Periods in The Lagoon Nebula\label{tab:data}}
\tablewidth{0pt}
\tablehead{\colhead{Column} & \colhead{Contents}}
\startdata
catnum & \cellp{IAU-compliant position-based catalog number -- also appears in \citet{lagooncatalog} unless incat=N}\\
ra & \cellp{right ascension (J2000)}\\
dec & \cellp{declination (J2000)}\\
diskcode & \cellp{disk code from \citet{lagooncatalog}; $\geq$3 suggests disk based on
available information. 0 = no data past \ks; 1 = no IR excess; 1.1 = no
significant excess, but all IRAC bands offset slightly above the photosphere
relative to \ks; 1.2 = as 1.1, but only IRAC-3 and IRAC-4 offset; 2 = probably
no excess; 3 = significant excess in the longest-wavelength band only; 3.1 =
excess significant in one band, with the adjacent shorter-wavelength band also
offset; 4 = likely excess; 5 = unambiguous excess}\\
bestk & \cellp{best available \ks, in Vega magnitudes}\\
obsv & \cellp{observed $V$, when available, in Vega magnitudes}\\
bestvk & \cellp{best available observed \vmk, in Vega magnitudes}\\
bestderedvk & \cellp{best available dereddened \vmkz, in Vega magnitudes}\\
p1 & \cellp{period in days, taken to be the rotation period of the star when available}\\
p2 & \cellp{second period, in days}\\
p3 & \cellp{third period, in days}\\
p4 & \cellp{fourth period, in days}\\
ztfconf & \cellp{ZTF confirmation of the period derived from K2: y = ZTF data confirm the K2 period; ? = ZTF data possibly confirm it; blank otherwise}\\
hendp & \cellp{literature period in days from \citet{2012ApJ...747...51H} (when available)}\\
ordenp & \cellp{literature period in days from \citet{2022MNRAS.517.6191O} (when available)}\\
dd & \cellp{if set, light curve is ``double dip'' type (two dips per cycle)}\\
ddmoving & \cellp{if set, light curve is ``moving double dip'' type (two dips per cycle that move with respect to each other)}\\
shch & \cellp{if set, light curve is ``shape changer'' type (shape changes over campaign)}\\
batwing & \cellp{if set, light curve is ``batwing'' type}\\
beater & \cellp{if set, light curve is ``beater'' type (beating apparent in light curve)}\\
complexpeak & \cellp{if set, light curve is ``complex peak'' type (structure seen in periodogram peak[s])}\\
resolvedclose & \cellp{if set, light curve is ``resolved close'' type  (peaks close in periodogram)}\\
resolveddistant & \cellp{if set, light curve is ``resolved distant'' type (peaks distant in periodogram)}\\
pulsator & \cellp{if set, light curve is ``pulsator'' type (many short-P peaks)}\\
disky & \cellp{if set, light curve variability resembles those from stars with disks}\\
dipper & \cellp{if set, light curve is ``dipper'' type}\\
burster & \cellp{if set, light curve is ``burster'' type}\\
binary & \cellp{if set, light curve is ``binary'' type -- e.g., suggests binarity}\\
iffy & \cellp{if set, period is ``iffy'' (uncertain, or may not be rotation)}\\
incat & \cellp{set to N if the star does not appear in the table in \citet{lagooncatalog} (see text)}\\
notes & \cellp{notes on object's color, light curve, and/or period}\\
\enddata
\tablecomments{For the light curve type columns (dd through iffy),
x = yes and ? = possibly; in the binary column, x? = probably.}
\end{deluxetable}

\subsection{Comparison to the Literature}

\begin{figure}[htb!]
\epsscale{1.0}
\plotone{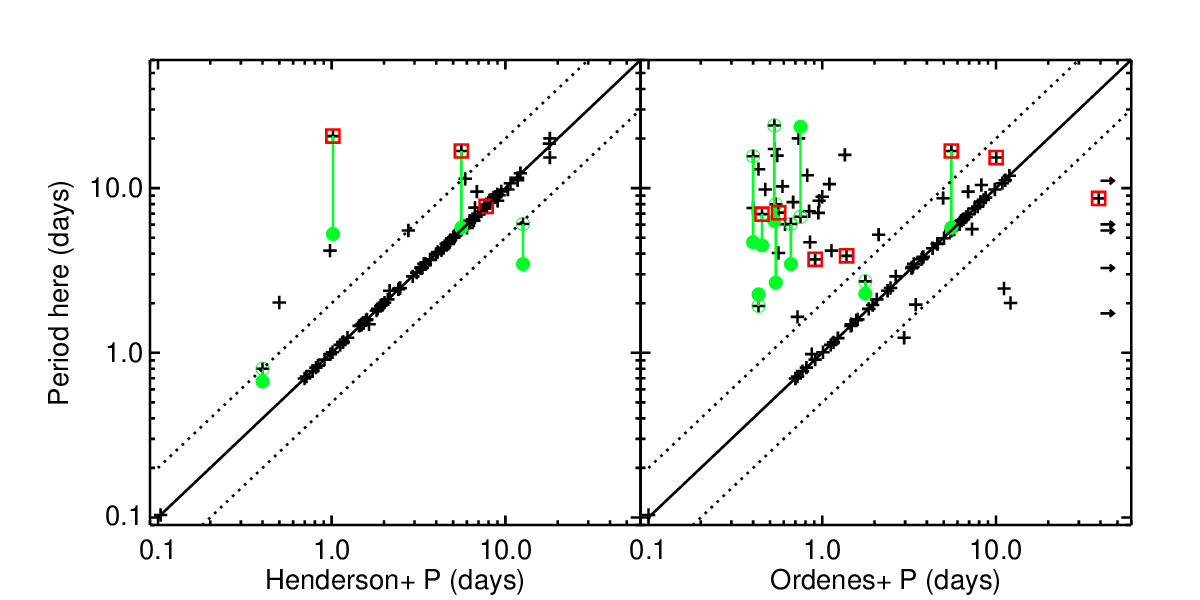}
\caption{Comparison of periods obtained here ($y$-axis) to those in
the  literature ($x$-axis), \citet{2012ApJ...747...51H} on the left
and \citet{2022MNRAS.517.6191O} on the right. The solid line is the
1:1 match, and dashed lines are a factor of 2 off. In general, the
periods we report here are well-matched to those in the literature.
Open green circles surround targets where the match to the first
period we report is significantly different than the reported period,
and they are linked with a solid green line to a filled-in circle at
the location of the second period we report for that target, where
available. In most cases, this second period moves the period reported
here closer to those reported in the literature for that target. 
Additional red squares indicate periods that were tagged uncertain
in our analysis (and are identified as such in our final catalog);
they are occasionally good matches to the literature values. 
(There are 5 periods from \citet{2022MNRAS.517.6191O} that are
$>$100d, and would be off  to the right here, represented by the
right-pointing arrows.)}
\label{fig:comparep}
\end{figure}

Figure~\ref{fig:comparep} compares the periods obtained here to those
found in the literature. In general, there is good agreement.

\begin{figure}[htb!]
\epsscale{1.0}
\plotone{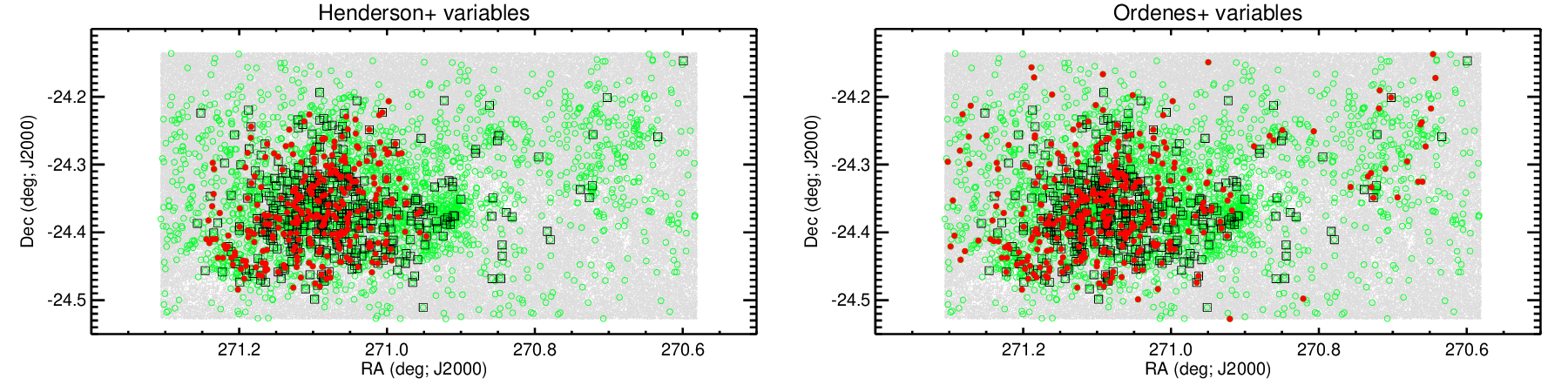}
\caption{Location on the sky of stars from 
\citet{2012ApJ...747...51H} on the left and
\citet{2022MNRAS.517.6191O} on the right, in red dots plotted on top
of the whole catalog in small grey dots from \citet{lagooncatalog},
with open green circles around the YSOs from \citet{lagooncatalog}. 
Additional black squares are the stars with K2 light curves presented
in this paper.  \citet{2012ApJ...747...51H} covers a smaller area
than  \citet{2022MNRAS.517.6191O}.}
\label{fig:wheresky}
\end{figure}

\citet{2012ApJ...747...51H} monitored the region in $I_{\rm C}$ 
with observing runs scattered over 6 observing seasons. 
We merged their catalog with ours in \citet{lagooncatalog}.
They have 100 targets in common with us and those targets are shown
on the sky in Fig.~\ref{fig:wheresky}; their field of view was smaller
than our field. 
Out of those stars in common, 87 match within 10\%, and an additional
6 match within $\sim$50\%. For several of the targets
that are mismatched, the periods we find are a factor of two off from
that in the literature.  For four targets, we report a second period
for that star, and in three of the four cases, this second period is
closer to the 1:1 line than the first period. In three cases, we have
noted the periods we report as perhaps uncertain, but one of those 
matches the literature value very well.

\citet{2022MNRAS.517.6191O} analyzed \ks\ light curves from the VISTA
Variables in the V\'ia L\'actea extension
\cite[VVVX;][]{2012AA...537A.107S,2017yCat.2348....0M}.   Cross-matching
their catalog with ours \citep[see ][]{lagooncatalog} yields 99
targets in common (see Fig.~\ref{fig:wheresky}). For 50 targets, the
periods agree within 10\%, with an additional 8 matching within 60\%. 
\citet{2022MNRAS.517.6191O} report periods for five targets to be
$>$100 d, which cannot be recovered given the $\sim$70 day K2
baseline.  For the remaining 44 discrepant sources, 27 ($\sim$60\%)
have K2 light curves characteristic of circumstellar disks, where
accretion or extinction variability can be seen on top of underlying
rotation. This intrinsic disk-related activity, combined with standard
ground-based observational constraints (day/night gaps, windowing
functions with short campaigns stretching over seasons, and
weather/seeing), particularly complicates period retrieval for active
young stars.  Additionally, six of the mismatches (and one of the
matches) have such strong stochastic contributions even in the K2
light curves that we flagged their periods as uncertain. In a few
cases where we report a second period, the second period is closer to
the 1:1 line than the first period.  We note also that several of the
particularly poor mismatches have literature periods that cluster
near $\sim$0.8-1~d, suggesting possible daily aliasing in the
ground-based data. (This clump includes four of the six targets
flagged with uncertain K2 periods.) 

Other contributing factors could be the  
light curves as observed themselves. The VVVX light
curves are on the order of 100 epochs spread over 12 years (median
time step $\sim$5 d), to be
compared with our $\sim$3270 epochs over $\sim$70 days (typical 
time step $\sim$30 min); the stars 
are likely to change more over years than over days, and sparse
light curves make it harder to find periods than dense light curves.
Moreover, there could be physical changes in the dominant
variability source between epochs; \citet{2022MNRAS.517.6191O} 
monitored in the NIR and K2 worked in the optical, so if there
are small changes in the disk between observation epochs, it could
lead to changes in light curve amplitude between optical and \ks. 
Perhaps related to this concept, interestingly, for two stars
(J180405.76-241455.0 and J180413.77-242405.1), we retrieve two
different periods. One of those periods  is retrieved (in the optical)
by \citet{2012ApJ...747...51H} and the other  (in the NIR) by
\citet{2022MNRAS.517.6191O}.  (See Table~\ref{tab:data} for
specifics.)

We conclude that our periods are at least as accurate as
those previously reported in the literature.

\section{Period Distributions}
\label{sec:pdistrib}

\subsection{A First Look}
\label{sec:firstpvk}

\begin{figure}[htb!]
\epsscale{1.0}
\plotone{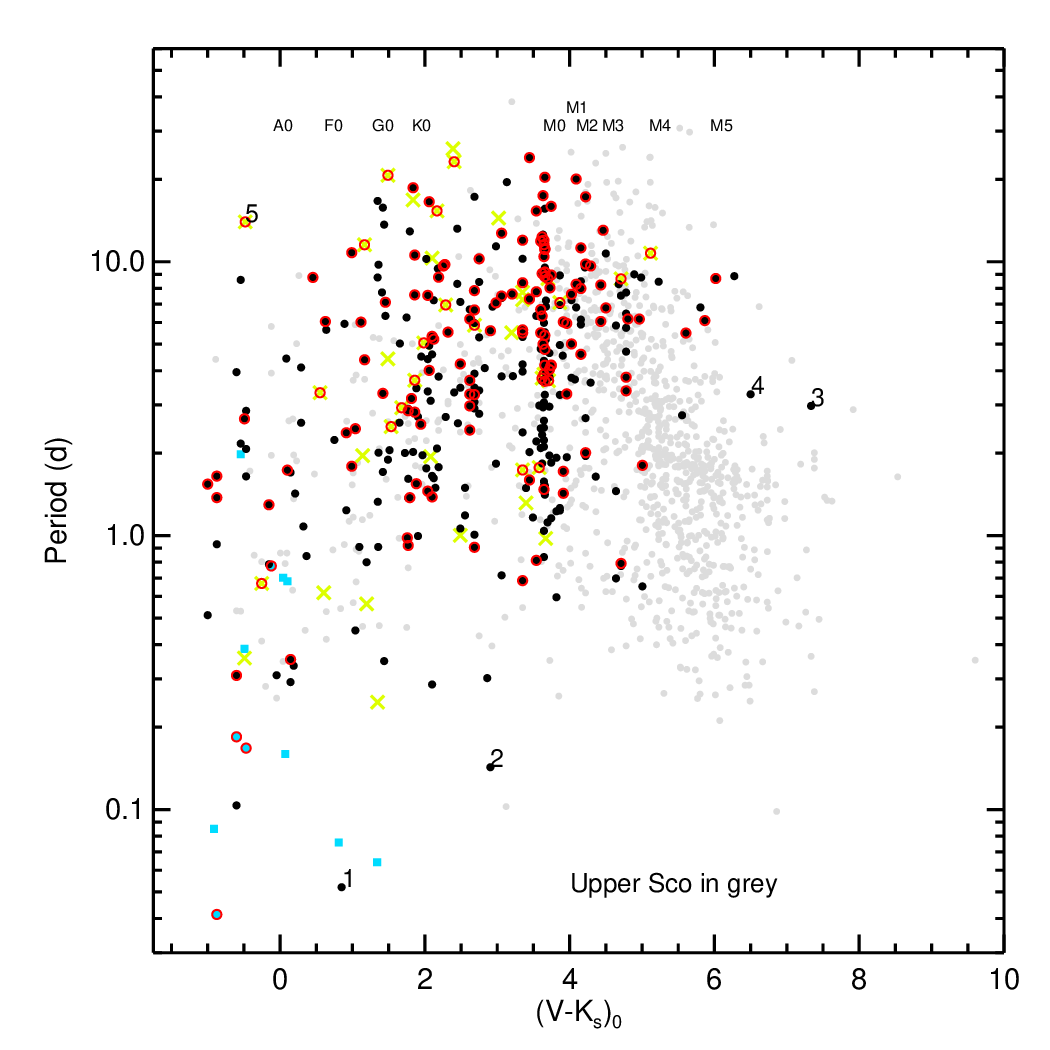}
\caption{The distribution of rotation rates (in days) as a function of
$(V-K_s)_0$ color as a proxy for mass in the Lagoon Nebula.
Approximate spectral types as a function of color are given along the
top of the plot. Cyan squares indicate   stars whose periods we
suspect may be pulsation (perhaps not rotation). Chartreuse $\times$
denote stars whose period may be less reliable than other stars here.
An additional red circle around a point denotes an IR excess (a likely
circumstellar disk). Grey points in the background come from Upper
Scorpius \citep{usco}, and are there to guide the eye.  Outliers are
numbered; see the text. The overall structure of the distribution
roughly matches that from USco -- mid-K has the slowest rotators and
both higher and lower masses rotate more quickly. See the text for
more details.}
\label{fig:lagoon2}
\end{figure}

Figure~\ref{fig:lagoon2} presents the first look at the distribution of
rotation rates as a function of color (a proxy for mass) in the
Lagoon Nebula. In this figure, the distribution of rotation rates
found in Upper Scorpius \citep[USco;][]{usco} is provided for reference, though
a more complete comparison to other clusters will be provided below. USco,
at $\sim$8 Myr, is slightly older than the $\sim$1-3 Myr Lagoon. USco is
significantly closer, so the population for which we have periods is
richer and reaches fainter masses.  The distributions are
roughly comparable -- the slowest rotators in the ensemble are near mid-K, 
\vmkz$\sim$2.5 to 3; the periodic A and F stars are rotating faster
than the mid-Ks; the mid-M stars are also rotating faster than the mid Ks. 

Figure~\ref{fig:lagoon2} requires some caveats. Due to  the reddening
issues discussed above, the \vmkz\ colors appear quantized,
particularly near \vmkz$\sim$3.5, because the stars are corrected to
the expected color based on spectral type, and spectral types are
quantized.  Figure~\ref{fig:lagoon2} has stars flagged where the
periods may not be as reliable as others here due to noisy data,
whether that be inherent noise in the light curve or `noise' from
accretion events; these are also called out in  Table~\ref{tab:data}.
These stars are often but not always outliers. These dubious periods
are omitted from later plots.

Figure~\ref{fig:lagoon2} also identifies stars for which periods may
be due to pulsation.  These pulsators are identified simply by the fact
that the periods are very fast (typically fractions of hours to a few
hours) and usually there are many viable peaks in the periodogram
(often $>$4 though we only record up to 4).  For the 12 possible
pulsators identified here, we report 4 periods for 9 of them, but
only 3 periods for 1, and only 2 periods for 2; the periods are fast
enough that we retained their tentative classification as pulsators. 
Experience with other clusters \citep[][and references therein]{usco}
suggests that, for these stars so-identified as pulsators, the
strongest peak in the power spectrum may be related to the rotation
period, in that for most of the stars flagged as possible pulsators, the
period corresponding to strongest peak places them in a location in
this diagram consistent with the rest of the rotation distribution. 

There are some remaining outliers of note called out in 
Figure~\ref{fig:lagoon2}, which we now briefly discuss. 

The star marked 1 is J180421.68-241547.0, and it has two periods, one
very fast (as shown), and one $\sim$4.3 d. If we were to move this
star to near 4 days in this plot, it would not be an outlier.  The
very fast peak is stronger in the periodogram than the 4 day peak,
suggesting we should keep the faster period as the primary period.
There is the very real question of whether or not this period should
be identified with this star, or if source confusion is important
here.  The ZTF light curve for this source is not publicly 
accessible, but the ASAS-SN \citep{asassn1,asassnvar23} light curve is
available, and also returns the fast period. Moreover, the ASAS-SN 
light curve's power spectrum does not have a peak consistent with the
slower period at all. Thus, we have retained this star and this period
here, but concede that it may be pulsation and not rotation.

The star marked 2 is J180325.93-242806.0 is a likely contact binary.
We know from the Pleiades \citep{ple3k2} that binaries are often 
fast-rotating outliers. 

There are two stars that are very red.  The star marked 3 is
J180358.81-242555.8 and while its rotation doesn't necessarily make it
an outlier, its color does. The SED is well-behaved (implying that the
$V$ estimate is good), and  there are two \ks\ measurements, from
2MASS and UKIDSS, very close to each other. Its placement in the
$JHK_s$ diagram suggests that there is IR excess that could be
affecting $K_s$, which  may explain its placement in \vmkz.  The star
marked 4 is J180405.41-242232.7, and again, its rotation  doesn't
particularly make it an outlier, but its color suggests extreme
redness. There is some scatter in the SED, but the  $V$ estimate is
probably accurate; the placement in $JHK_s$ suggests that the $K_s$ is
photospheric. \citet{2012ApJ...747...51H} reports the same period 
that we do for this star, suggesting the period is ok.  We have no
clear explanation for the location of this star.

The star marked 5 is J180450.61-242541.8 and the period we have for 
it, $\sim$14 d, suggests it is very slowly rotating given its color.
This star has an unambiguous IR excess (disk) and its light curve
suggests dust occultations. There are variations on multiple
timescales. The longer period that we have tied to this star seems
repeatable/reliable, but perhaps it is not stellar rotation. There is
a period near $\sim$0.4 d that one can kind of identify by eye, but
it is not reliable over the entire K2 campaign. Knowing that, one can
look for a similar peak in the ZTF light curve, and it phases well,
but one would not have selected that peak from ZTF without that prior
knowledge. We have left this star at the $\sim$14 d period as a
result, but it is possible that it should be tagged with a faster
period.

\subsection{In Context with Other Young Clusters}
\label{sec:otherclusters}

\begin{figure}[htb!]
\epsscale{1.0}
\plotone{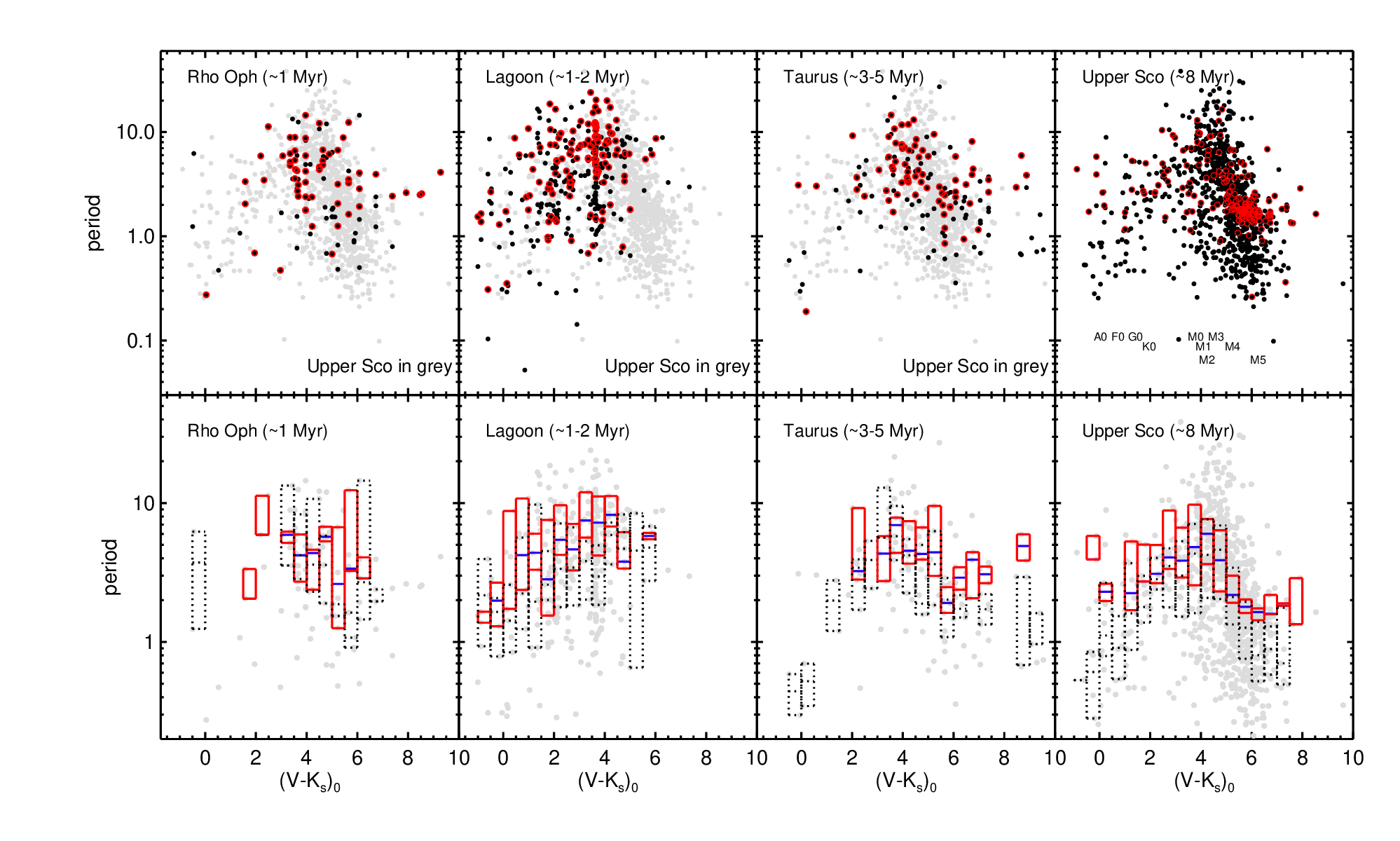}
\caption{The distribution of rotation rates (in days) as a function of
$(V-K_s)_0$ color as a proxy for mass in four clusters: $\rho$ Oph
($\sim$1 Myr),  Lagoon ($\sim$1-2 Myr), Taurus ($\sim$3-5 Myr), and 
USco ($\sim$8 Myr), as indicated; all of these are based on K2
data.  In the top row, an additional red circle denotes an IR excess
(at any band), \eg, a disk. In the first three panels of the top row,
the data from USco are plotted in grey to guide the eye. Pulsators
and low-confidence members and periods are omitted from all panels.
Approximate spectral types are indicated in the last panel of the top
row. The bottom row is the same four clusters in the same order (note
the $y$-axis is slightly reduced compared to the top row for
clarity), but now the grey points are the corresponding cluster's
points, and the boxes show the median, first, and 3rd quartile of the 
distribution within that color range. The red solid lines are disked
stars (with the medians blue) and the black dotted lines are the disk-free stars. 
The new Lagoon data presented here represents a substantial addition
to the known rotation rates for 1-5 Myr stars. 
The overall structure of the $P$ vs.\ color distribution 
is similar across the four clusters. Disked rotation rates are 
slightly slower than that for disk-free. See the text for more details.}
\label{fig:multicluster4}
\end{figure}

Figure~\ref{fig:multicluster4} puts these Lagoon results in context
with three other of our youngest clusters with K2 data, analyzed in
exactly the same way: $\rho$ Oph \citep[$\sim$1 Myr;][]{usco}, Taurus
\citep[$\sim$3-5 Myr;][]{taurusk2}, and USco \citep[$\sim$8
Myr;][]{usco}. The top row of Figure~\ref{fig:multicluster4} has the
clearest, best possible distributions and the bottom row has the
medians, 25th, and 75th percentile of the distributions shown in 
\vmkz\ color bins.  

The first thing to notice is the number of points in each panel in the
first row. The panel for USco is very well-populated -- there are
nearly 1000 stars in that plot, to be compared with $\sim$200 in
Taurus, and $\sim$100 in $\rho$ Oph.  The 325 (high confidence,
unlikely pulsator) stars that we can now add from the Lagoon represent a
substantial addition to the distribution of K2 stellar rotation rates
available for the $\sim$1-5 Myr population. 

\begin{figure}[htb!]
\epsscale{0.5}
\plotone{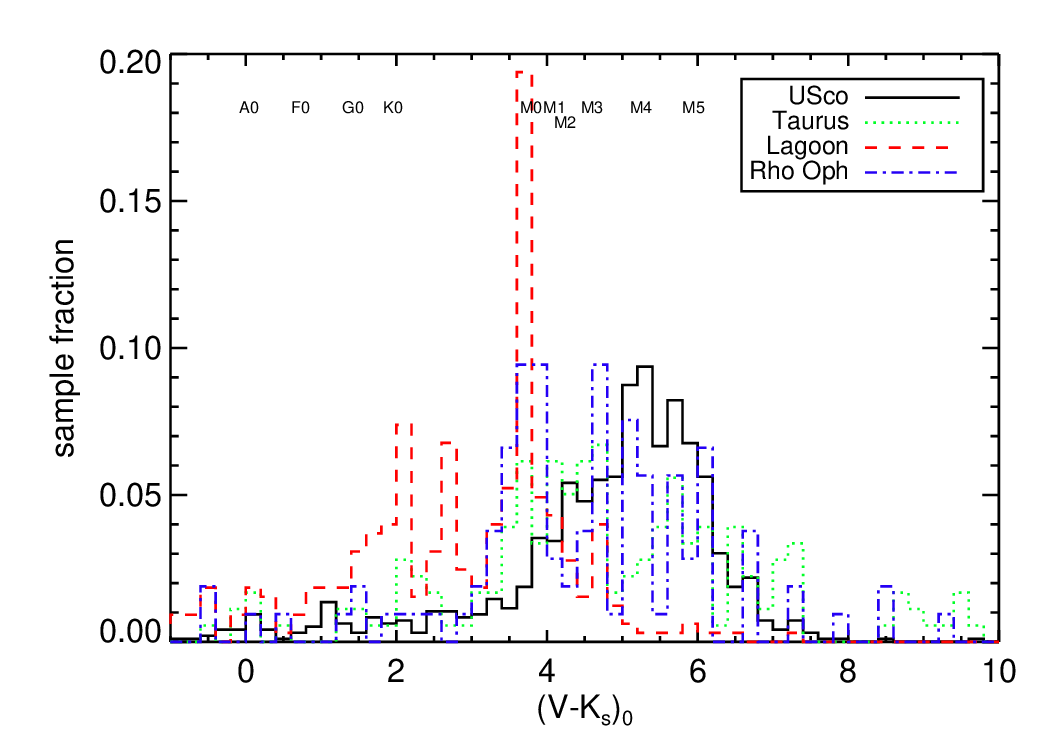}
\caption{The distribution of $(V-K_s)_0$ colors for the stars
with periods (those that appear in Fig.~\ref{fig:multicluster4})
in four clusters: $\rho$ Oph ($\sim$1 Myr; blue dash-dot),  
Lagoon ($\sim$1-2 Myr; red dashed), 
Taurus ($\sim$3-5 Myr; green dotted), and 
USco ($\sim$8 Myr; black solid), as indicated. Approximate
spectral types are shown across the top of the plot. It is again
clear that data from the Lagoon do not reach as far into the M 
stars as the other clusters.}
\label{fig:colorhisto}
\end{figure}

Unavoidably, $\rho$ Oph, Taurus, and USco are only about 
$\sim$140 pc away, whereas the Lagoon is much further 
($\sim$1250 pc). Therefore, we are unable to probe very
deeply into the lowest mass stars in the Lagoon, certainly 
in comparison to these other three clusters.  
Figure~\ref{fig:colorhisto} shows histograms of $(V-K_s)_0$
for the four clusters in Fig.~\ref{fig:multicluster4}. 
It is clear that the Lagoon data cut off in the early Ms 
whereas the rest of the clusters reach mid-Ms. 

Another feature apparent in the rotation rate distribution is that
despite selection effects imposed by fewer stars and the lower mass
cutoff in the Lagoon, there is an overall similar structure in all
four of these clusters -- the slowest rotators are found near
\vmkz$\sim$3.5-4, with faster rotators both bluer and redder than
that. Because there are many more stars in USco, that distribution is
the most structured of the four, but based on the medians in the
bottom row of Fig.~\ref{fig:multicluster4},  Taurus has that structure
clearly in place as well. It is less evident in $\rho$ Oph because
there are fewer stars.  The structure is even less obvious in the
Lagoon,  as a result of the many fewer mid- to late-M stars.

For all of these clusters, the light curves are space-based, so the
periods are either known very precisely, typically to 0.001 d, or are 
incorrect by a significant margin (factor of 2 aliasing, simply wrong
due to source confusion, or simply wrong due to stochastic noise most
often from accretion/dust occultation in the light curves). 
Additionally, for all four of these clusters, reddening plays an
important role. We have endeavored to mitigate this (see reddening
discussions in the corresponding papers for the other clusters). The
quantization in \vmkz\ that comes from dereddening according to spectral
types can be seen in at least three of the four clusters. However,
there is an unavoidable effective smearing in the $x$-direction that
must still be present in these plots, particularly in the first three
relatively sparse and youngest (dustiest) clusters. It is encouraging
that the overall structure nonetheless seems to still emerge in all of
these clusters.  It is, however, clearly noisy, particularly in the
two youngest clusters, likely exacerbated by the relatively small
sample size, variable reddening, and limited depth in the Lagoon.

\subsection{Influence of Disks}
\label{sec:disks}

Figure~\ref{fig:multicluster4} also distinguishes between the disked
and disk-free stars. For the Lagoon, there are a total of 
124 disked stars and 201 disk-free stars with reliable rotation rates. 

USco has so many M stars that no sophisticated statistics are
required to detect the pile-up of disked M stars at $\sim$2d (also see
discussion in \citet{usco}).  It is obvious in the corresponding bins
of the USco median plot that the ensemble of disked M stars tends on
average to rotate more slowly than the disk-free M stars. This trend
continues in the other clusters as well, but the sample sizes are much
smaller, and so the significance is correspondingly weaker. The 75th
percentile ends of the boxes for the disked stars always extend above
that for the disk-free stars in USco \vmkz$\sim$1-4. In Taurus, among the
10 bins where there are both disked and disk-free stars, seven have
the 75th percentile end of the boxes for the disked stars extend above
that for the disk-free. For the Lagoon, there are 13 bins where there
are both disked and disk-free stars, and 9 have the 75th percentile
end of the boxes for the disked stars extend above that for the
disk-free stars. Among the Lagoon bins, the two bins that have the largest
discrepancy between disked and disk-free are also the best-populated
bins (\eg, should be the most reliable). 

We conclude that there is weak evidence that the disk-bearing stars
rotate, on average, more slowly than the disk-free stars,
particularly in the late K/early M regime, even for stars as young as
in the Lagoon.  Assessing this in any more detail is greatly
complicated by the relatively small sample sizes found in the youngest
three clusters, and the uncertainties of dereddening, as well as the fact
that stars  with disks often have complex light curves where periods
are  harder to extract; see next section.

\section{Other Light Curve Properties}
\label{sec:other}

\subsection{Dippers and Bursters and Other Disk-Influenced Light Curves}
\label{sec:diskylcs}

\citet{venuti2021} discussed the multi-color variability properties 
of the light curves, particularly of the disked population, so we
make only limited comments here.

Of the 16 stars we have identified as likely bursters, all but one of them 
are also (independently) identified as having an IR excess (circumstellar disk). 
However, that one apparent disk-free star (J180407.27-242527.6) has very
few points longer than 2 \mum, and in the context of the rest of the 
full ensemble of stars considered in \citet{lagooncatalog}, any evidence
for (or against) an IR excess is very weak. 

Of the 39 stars we have identified as likely dippers, similarly most of them
are also (independently) identified as having an IR excess (circumstellar disk). 
Of the 6 that are not tagged as having a disk, 1 has no data past 2 \mum, 2 
are subject to the systematic IRAC offset (see above), and the remaining
3 similarly have simply weak (or ambiguous) evidence for an IR excess. 

We have tagged 91 stars as generally having light curve patterns 
consistent with other stars that have disks; all but 11 independently
have IR excesses. 

\begin{figure}[htb!]
\epsscale{1.0}
\plotone{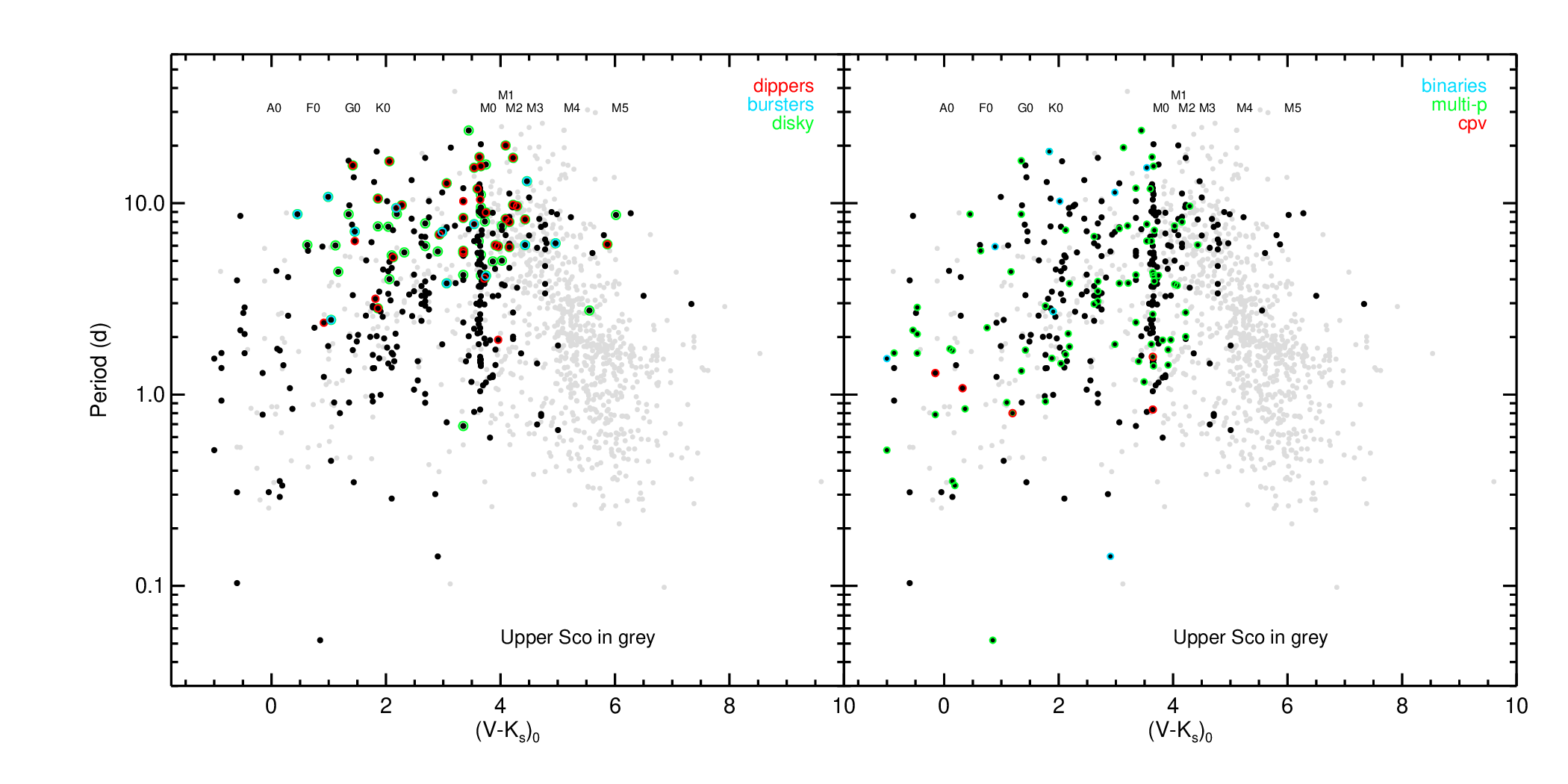}
\caption{Another view of $P$ vs.\ $(V-K_s)_0$ color, as before (with
uncertain (`iffy') and pulsating periods removed), but with dippers
(red), bursters (blue), and with characteristics similar to that of
stars having disks (`disky'; green) light curves highlighted on the
left,  and binaries (blue), multi-periodic stars (green), and CPVs
(red; see Sec.~\ref{sec:cpvs}) highlighted on the right. Stars with 
indications of having a
disk (light curves suggesting presence of a disk, e.g., really
anything in the left panel) tend towards longer periods. Binaries in
other clusters tend to be the fast outliers, but apparently not always
here; just one is a fast outlier. CPVs are biased towards fast 
periods but are not M stars per se. Multi-periodic stars are found 
throughout the brighter stars; presumably the fainter stars have light
curves that are too noisy to easily reveal multiple periods. See the
text for more details.}
\label{fig:highlights2}
\end{figure}

Figure~\ref{fig:highlights2} shows that these objects with 
light curves suggestive of disks have a vague tendency towards 
longer periods, consistent with above when we considered solely 
those with IR excesses.

\subsection{Multi-Periodic Stars \& Binaries}
\label{sec:binaries}

Photometric binary stars in the Pleiades tend to be fast-rotating
outliers \citep{ple3k2}. It is, however, far easier to identify
photometric binaries in the Pleiades than it is in the Lagoon or
other very young clusters where the stars are not only subject to
reddening but also have not yet settled into a narrow isochrone at the
main sequence.  We can identify nine eclipsing binaries, or possible
eclipsing binary candidates, in our Lagoon sample. In several cases,
the period we report here is obtained by omitting the eclipses and
searching for periods in what, presumably, is the remaining
photospheric signal. (These cases are noted in the table.) These
binaries do not particularly seem to be rotating quickly, except for
the one outlier noted in Sec.~\ref{sec:firstpvk} above. 

In USco \citep{usco}, UCL/LCC \citep{ucllcc}, the Pleiades
\citep{ple2k2}, and Praesepe \citep{k2praesepe}, multi-periodic M
stars were often likely binary M stars; higher-mass multi-periodic
stars could suggest binarity or could be multiple spot/spot groups
differentially rotating on a single star. Here in the Lagoon, as seen
in Figure~\ref{fig:highlights2}, there are few multi-periodic stars 
with \vmkz$\gtrsim$4.5, presumably mostly due to the fact that the
stars are too faint for multiple periods to be easily extracted from
the light curve. Bluer than that, however, there are multi-periodic
stars scattered through the distribution.  In upcoming
work, we analyze speckle photometry to further tease out the effect of
multiplicity on periodic and other variability types.

\subsection{CPVs}
\label{sec:cpvs}

\begin{figure}[htb!]
\epsscale{1.0}
\plotone{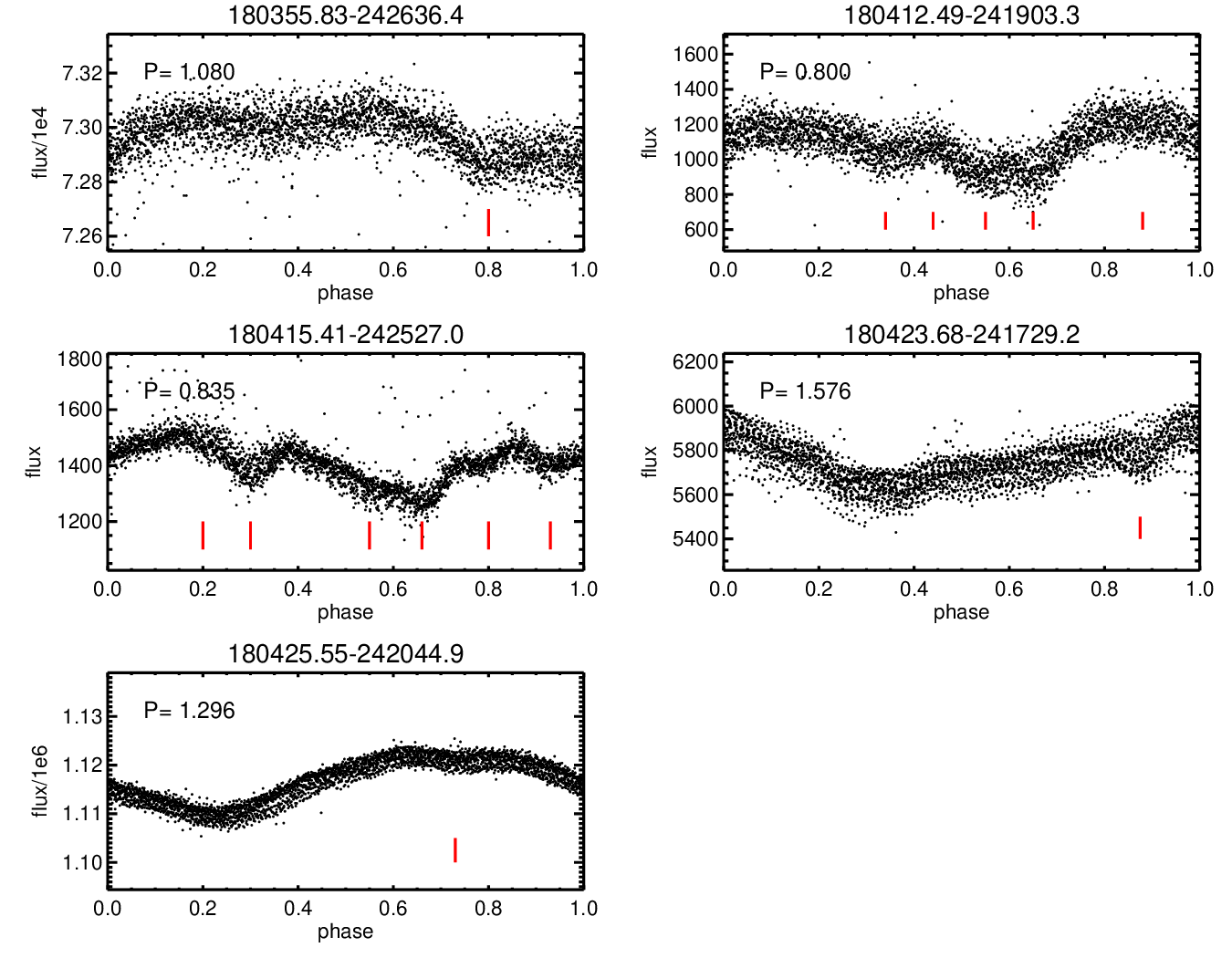}
\caption{Phased light curves for the CPVs discussed in
Sec.~\ref{sec:cpvs}, in the same order as they appear in the text. Red
hash marks indicate the location  of the sharp-edged dips. The last
one is the earliest spectral type (a B star)  and is least like a CPV
in shape. See the text for more details.}
\label{fig:cpvs}
\end{figure}

Based on K2 data, we identified a class of light curves that we 
suspect are a result of orbiting clouds of material 
\citep{2017AJ....153..152S,2018AJ....155...63S,2021AJ....161...60S}. 
In  \citet{2017AJ....153..152S}, the objects were defined as being
all young, disk-free, rapidly rotating M stars, and placed into categories
dubbed `scallop shells', `persistent flux-dip', or `transient narrow dip.'
Scallop shells were required to have periods less than 0.65 days. 

As the hunt for these kinds of objects expanded
\citep[e.g.,][]{2024AJ....167...38B,2025ApJ...988L...3B,2026MNRAS.549ag837D}, 
the nomenclature changed to ``complex periodic variables (CPVs)'' 
because the light curve shapes -- with sharp-edged structure -- resulted in 
power spectra with many peaks, most of which are harmonics of the main period.
The definition expanded slightly to no longer require that the stars
have $P<$0.65 d or strictly be M stars, though the basic 
physical interpretation of orbiting clouds of material has persisted. 

There are five stars in the Lagoon that bear hallmarks of these kinds
of light curves, though some are confusing, and none are M stars. 
Since all of these stars are earlier types (three far earlier type)
than the `classical' requirement of an M star, perhaps they are not CPVs
at all. Or, it's possible that these stars are multiples of some sort
and the companion is the CPV. If
this is the case, the high mass star would contribute little or no
variability to the light curve, and the CPV portion of the light curve
would come from the later-type companion. For the four of
these that have Gaia data, Gaia does not indicate (via RUWE) that they 
are multiples (though these stars may well be too far away for RUWE to
be helpful). Moreover, the amplitude of the variability suggests that 
companions can't be responsible.

We now discuss each of these stars.  They are plotted in the $P$ vs.\
\vmkz\ diagram in  Fig.~\ref{fig:highlights2}, where they clearly skew
towards faster rotation in general, but not all as fast as 0.65 d.
The  phased light curves appear in Fig.~\ref{fig:cpvs}, in the same
order in which they appear here, with the angular dips indicated.

{\bf J180355.83-242636.4:  } This star is a A3.75 \citep{venuti2024},
so the  \vmkz\ is 0.322. It has been identified by five different
studies as a YSO,  largely but not exclusively on the basis of X-rays.
It does not have compelling evidence for a circumstellar disk --
IRAC-3 is significantly above the photosphere, and there is a MIPS-24
limit near the photosphere, suggesting that the IRAC-3 point is
erroneous. The light curve has many low drops below the continuum
(cropped out of  Fig.~\ref{fig:cpvs}), as if they were short
eclipses, but we could find no period that would synch them. There is
clear angular structure in the main portion of the light curve
resulting in many harmonic peaks in the power spectrum, resulting in
this star being classified as a CPV. However, its main rotation period
is 1.080 d, longer than the `classical' definition.  This star is 
bright, and in a region of sky with high surface brightness, such 
that we do not have a lot of optical data for it (including no Gaia
data), so it is difficult to constrain based on ancillary data 
whether or not it has a companion.

{\bf J180412.49-241903.3:  } \citet{venuti2024} has this star as F5.5,
meaning that its  \vmkz\ is 1.195. This star is unlikely to have an IR
excess, but the IR is systematically offset from a Rayleigh-Jeans
line  extended from \ks\ (see discussion in \citet{lagooncatalog}). It
is identified by seven different papers in the literature as a YSO.
This light curve is  a stereotypical CPV shape, albeit noisy, with
several angular dips per cycle; at 0.800 d, it is a little slow
compared to the `classical' definition, but as an F star, it is also
more massive. 

{\bf J180415.41-242527.0:  } This star has been identified by 10
different  articles as a YSO. \citet{venuti2024} classifies it as a
K7.5, so its \vmkz\ is 3.645.  It does not appear to have an IR
excess. This star also has a stereotypical CPV light curve shape,
with several sharp, angular dips per cycle. It, too, is a bit slow at
0.835 d. It has a few obvious flares during the observing campaign.
This is one  of the latest-type CPV stars we have identified, and thus
closest to the `classical'  definition. 

{\bf J180423.68-241729.2:  } This is also a K7.5 \citep{venuti2024}
and therefore has  \vmkz = 3.645; it is identified by 12 different
papers as a likely YSO. It does not appear to have an IR excess. At
1.576 d, its period is fairly long for a CPV. Its light curve is more
ambiguous than the other CPV candidates here, less obviously CPV. It
has one dip that is angular enough to produce several harmonic peaks
in the periodogram. 

{\bf J180425.55-242044.9: } This is the most confusing of the stars in
this list.  \citet{venuti2024} has it as a B8.75, and it appears to
have a circumstellar disk,  with a significant IR excess at least at 8
and 24 \mum\ if not 5.8 \mum\ as well.  It's been identified by 7
studies as a YSO.  It has a period of 1.296 d, so very slow for a CPV.
It does not have myriad peaks in the periodogram, but it does have a
somewhat angular dip in the phased light curve, which is why we
initially tentatively put it in this  class (initial classifications
of light curves are done independently of any other  information about
the object such as spectral type). Since it is a B star, has a disk,
and the period is so long, plus  the fact that its light curve shape
is the least stereotypically CPV of all of them here, all of this
suggests that perhaps it should not be in the CPV class --  maybe the
structure is an eclipse and/or it is a binary instead. However, it 
could also be of the family of high-mass analogs mentioned in 
\citet{2025ApJ...988L...3B}. 

From our work in other clusters, we expect a higher rate of these
kinds of CPVs in younger clusters -- $\sim$1\% of Pleiades stars,
$\sim$3\% of UCL/LCC, $\sim$3\% of USco, and $\sim$4\% of Taurus
stars are like this. Out of the best sample of stars with good
rotation periods and colors, there are 325 Lagoon stars, and if 4 or
5 of them are CPVs, that's just $\sim$1-2\%. However, the prior
definition required that CPV stars be M stars, and all of the prior
cluster data had many more M stars, so it is less surprising that the
sample fraction for the Lagoon is different.

\section{Summary}
\label{sec:summary}

In this paper, we have worked with 414 K2 light curves for young stars
in the  Lagoon Nebula, one of the youngest \citep[$\sim$1-2
Myr;][]{2019AA...623A.159P} clusters with space-based rotation rates
available.  We find 376 periods, 325 of which are unambigous periods
and likely rotation (as opposed to pulsation).  Since the cluster is 
at $\sim$1250 pc \citep{2019AA...623A.159P}, the faintest available
rotation rates only reach the earliest M stars.   The overall
structure is broadly consistent with expectations in that the  slowest
rotation rates are near mid-K, with both higher and lower masses
rotating faster. Out of the 325 highest quality rotation periods, 124
are disk-bearing and 201 are disk-free.  There is some weak evidence
that stars with disks may be rotating slower on average, but there are
simply not enough stars to  assert this with any confidence.
Multiperiodic stars are located throughout the distribution.  

There are a handful of CPV stars, which is not surprising because the
Lagoon is young, but what is surprising is that none of these CPV
stars are M stars -- they are all mid-K or earlier. It's possible that
the stars creating the CPV structure in the light curves are unseen
low-mass companions, or the stars we have identified here as CPVs are
not true CPVs, but something other than the M stars that have been 
previously used to define the CPV category.

\begin{acknowledgments}
\begin{itemize}
\item This research has made use of the NASA/ IPAC Infrared Science Archive (IRSA), which is operated by the Jet Propulsion Laboratory, California Institute of Technology, under contract with the National Aeronautics and Space Administration.
\item This paper includes data collected by the Kepler mission and obtained from the MAST data archive at the Space Telescope Science Institute (STScI). Funding for the Kepler mission is provided by the NASA Science Mission Directorate. STScI is operated by the Association of Universities for Research in Astronomy, Inc., under NASA contract NAS 5–26555.
\item ZTF is supported by the National Science Foundation and a collaboration including current partners Caltech, IPAC, the Oskar Klein Center at Stockholm University, the University of Maryland, the University of California at Berkeley, the University of Wisconsin at Milwaukee, University of Warwick, Ruhr University, Cornell University, Northwestern University and Drexel University. Operations are conducted by COO, IPAC, and UW.
\item This publication makes use of data products from the Wide-field Infrared Survey Explorer (WISE), which is a joint project of the University of California, Los Angeles, and the Jet Propulsion Laboratory/California Institute of Technology, funded by the National Aeronautics and Space Administration.
\item This work is based in part on archival data obtained with the Spitzer Space Telescope, which was operated by the Jet Propulsion Laboratory, California Institute of Technology under a contract with NASA.
\item This publication makes use of data products from the Two Micron All Sky Survey (2MASS), which is a joint project of the University of Massachusetts and the Infrared Processing and Analysis Center/California Institute of Technology, funded by the National Aeronautics and Space Administration and the National Science Foundation.
\item Herschel is an ESA space observatory with science instruments provided by European-led Principal Investigator consortia and with important participation from NASA.
\item This research made use of data products from the Midcourse Space Experiment (MSX). Processing of the data was funded by the Ballistic Missile Defense Organization with additional support from NASA Office of Space Science. 
\item The Pan-STARRS1 Surveys (PS1) and the PS1 public science archive have been made possible through contributions by the Institute for Astronomy, the University of Hawaii, the Pan-STARRS Project Office, the Max-Planck Society and its participating institutes, the Max Planck Institute for Astronomy, Heidelberg and the Max Planck Institute for Extraterrestrial Physics, Garching, The Johns Hopkins University, Durham University, the University of Edinburgh, the Queen's University Belfast, the Harvard-Smithsonian Center for Astrophysics, the Las Cumbres Observatory Global Telescope Network Incorporated, the National Central University of Taiwan, the Space Telescope Science Institute, the National Aeronautics and Space Administration under Grant No. NNX08AR22G issued through the Planetary Science Division of the NASA Science Mission Directorate, the National Science Foundation Grant No. AST–1238877, the University of Maryland, Eotvos Lorand University (ELTE), the Los Alamos National Laboratory, and the Gordon and Betty Moore Foundation.
\item This work is based in part on data obtained as part of the UKIRT Infrared Deep Sky Survey (UKIDSS).
\item This work has made use of data from the European Space Agency (ESA) mission
{\it Gaia} (\url{https://www.cosmos.esa.int/gaia}), processed by the {\it Gaia}
Data Processing and Analysis Consortium (DPAC,
\url{https://www.cosmos.esa.int/web/gaia/dpac/consortium}). Funding for the DPAC
has been provided by national institutions, in particular the institutions
participating in the {\it Gaia} Multilateral Agreement.
\item The VVV is based on data products from observations made with ESO
    Telescopes at the La Silla Paranal Observatory under programme ID 179.B-2002
\item The VPHAS catalogue is based on data products from observations made with ESO Telescopes at the La Silla Paranal Observatory under public survey programme ID, 177.D-3023
\item This work was partially supported by the National Aeronautics and Space Administration (NASA) under grant No. 80NSSC21K0633 issued through the NNH20ZDA001N Astrophysics Data Analysis Program (ADAP).
\item Generative AI tools were used in preparing this work. Gemini (Google) was used
to suggest rewording of individual paragraphs. Claude (Anthropic; Claude Opus 5.5),
used through the Claude desktop app (Cowork), was used to debug and cross-check the
IDL code used to assemble the photometry and colors, to check the machine-readable
version of Table~1 against the pipeline outputs, to make the notes in that table
consistent, and to draft or suggest wording for parts of the text (e.g., the
description of stars not in the companion catalog and of the table columns). All
AI-assisted text, code changes, and table contents were reviewed, verified, and
revised by the authors, who take full responsibility for the content of this paper.
\end{itemize}
Thanks for everything, Mom. Miss you.
\end{acknowledgments}

\begin{contribution}
Co-Author Cody reduced the light curves from K2. Co-Author
A.~Thanvantri worked on the PSF-fitted light curves, and Co-Author
S.~Thanvantri did an early analysis of the rotation periods. Co-Author
Venuti analyzed the light curves initially in \citet{venuti2021}. 
Author Rebull analyzed and classified the K2 light curves in exactly
the same way as other papers in her series of K2 rotation periods; she
combined it with the catalog from \citet{lagooncatalog}, then did the
remaining analysis and wrote the paper.  Co-authors Cody and Venuti
helped resolve discrepancies between the papers and commented on the
manuscript. 
\end{contribution}

\facility{IRSA}

\bibliography{newbiblio}{}
\bibliographystyle{aasjournalv7}

\end{document}